\documentclass[conference]{IEEEtran}

\usepackage{mymacros}

\begin{document}

\bstctlcite{IEEEexample:BSTcontrol}

\title{Application of Flexible Numerology to Blockage Mitigation in 5G-mmWave Networks}

\author{
\IEEEauthorblockN{Fadhil Firyaguna, Jacek Kibi\l{}da, Nicola Marchetti}
\IEEEauthorblockA{CONNECT Centre, Trinity College Dublin, Ireland \\
\{firyaguf, kibildj, nicola.marchetti\}@tcd.ie}
}

\maketitle

\begin{abstract} 
The 5G New Radio (NR) standard for wireless communications supports the millimetre-wave (mmWave) spectrum to yield unprecedented improvement of the access network capacity.
However, intermittent blockages in the mmWave signal may degrade the system performance and lead to the under-utilisation of the allocated resources.
To circumvent this problem, the transmission slot-time shall be adjusted according to the blockage condition, avoiding the resource under-utilisation.
In this paper, we propose that the 5G NR flexible numerology should be applied to adapt the slot-time in order to mitigate the blockage effects. We validate this claim by analysing the expected data rate of a mmWave system, under a range of blockage scenarios.
We show that different blockage scenarios may require different numerologies to produce best performance, and that the correct choice of numerology may improve this performance by as much as hundreds of Mbps. Our results carry insights important for the design of blockage-aware scheduling mechanisms for 5G.
\end{abstract}


%
\IEEEpeerreviewmaketitle

\section{Introduction}
The fifth-generation (5G) of mobile networks is being developed to boost the available mobile speeds to multi-Gbps, and, consequently, provide support for the increasing user traffic demands \cite{itu2018setting}.
To achieve this goal, 5G networks will use the wide bandwidths available in \ac{mmWave} frequencies.
The challenge is that ordinary objects (e.g., human bodies, furniture) that are \yale{transparent} to signals transmitted over microwave frequencies become blockages when the same signals are transmitted over \acp{mmWave}.

Blockages in \ac{mmWave} signal propagation are related to severe attenuation of the signal power (in certain cases, the blockage may add as much as \unit[40]{dB} of attenuation \cite{maccartney2018rapid}), which can lead to radio link failures and consequent disconnection in the communication.
This issue has mostly been addressed in the literature to date by deployment strategies that allow the network to exploit spatial macro-diversity, i.e., increasing the communication robustness by enabling the user to receive a signal from distinct points in space.
These deployment strategies include: reflective surfaces \cite{narayanan2017coverage}, relay nodes \cite{kim2017relay}, dense networks \cite{bai2015coverage}, ceiling-mounted \acp{AP} \cite{firyaguna2017coverage}, and movable \acp{AP}  \cite{gapeyenko2018effects} that can position themselves in a way that increases the likelihood of having an \ac{AP} operating in \ac{LOS}.
In order to achieve spatial macro-diversity, the \ac{MAC} layer mechanisms should properly coordinate the network nodes and allocate the transmission resources (e.g., time, frequency, space) according to the blockage condition.
Yet, as we show in our numerical results, the intermittency of blockage events may cause system performance degradation and lead to resource under-utilisation if a fixed \ac{TTI} is considered.

\begin{figure}[b]
\centering
\includegraphics[width=.95\linewidth]{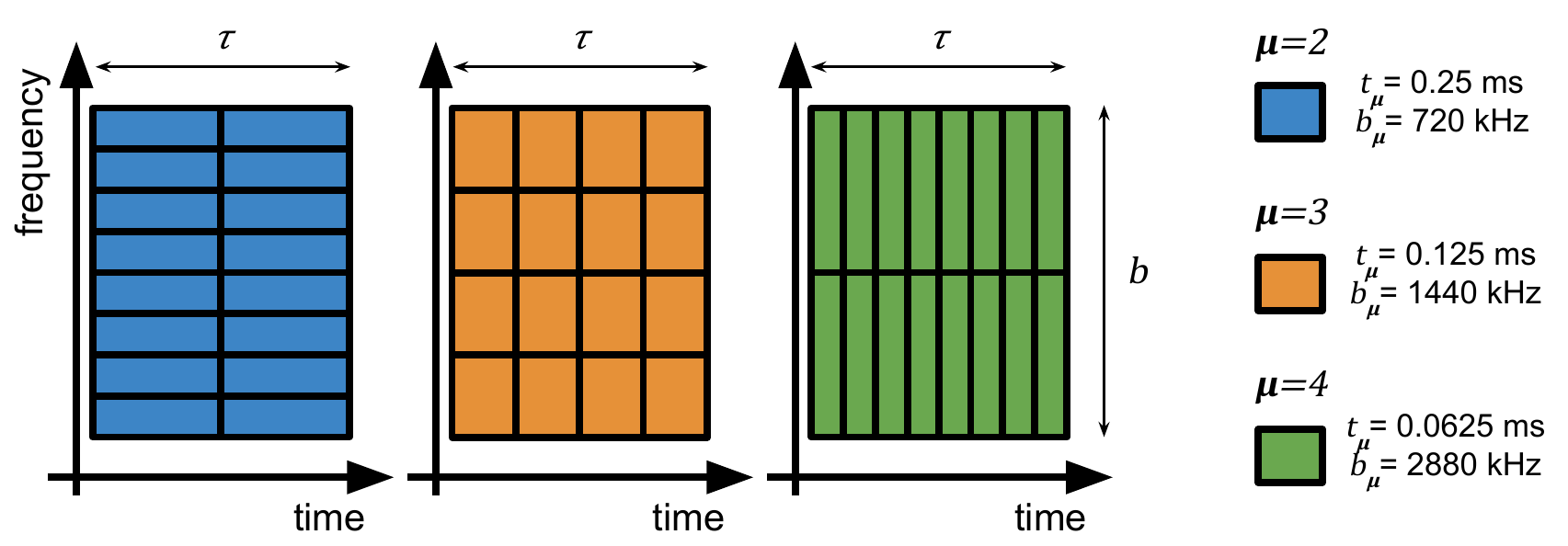}
\caption{Exemplary application of different numerologies in a 5G frame. The resource blocks are allocated for \unit[$\tau$]{ms} and within a bandwidth $b$ .}
\label{fig:example_scheduling_5G}
\end{figure}

In this paper, we consider adaptable transmission times for blockage mitigation, using the flexible \ac{TTI} proposed for the 5G \ac{NR}.
The ability of \ac{TTI} adjustment is enabled by the 5G \ac{NR} access technology. It works with a flexible \ac{OFDM} transmission frame system, in which the configuration of \ac{TTI}, i.e., \ac{SCS} and \ac{CP}, is flexible, as illustrated in Figure \ref{fig:example_scheduling_5G}. According to the 5G terminology \cite{5gnr_38211}, such configuration is referred to as the \textit{flexible numerology} and the supported numerologies are listed in Table \ref{tab:resource_type}. Originally flexible numerology was introduced to enable service-level differentiation, i.e., network slicing for different 5G use cases \cite{shafi2017tutorial}.

Herein, we propose an alternative application for flexible numerology. Our claim is that different numerologies will fare better under blockage conditions, and hence may be used to improve the mmWave user performance. We verify our claim by analysing the mmWave link performance using the numerologies available for 5G mmWave systems, under a range of blockage scenarios, defined and empirically-validated in \cite{yoo2017channel}. Our results show that there is a trade-off between the high transmission efficiency, achieved with longer \acp{TTI}, and the high probability of LOS transmission, achieved using shorter \acp{TTI}. In consequence, the same numerology used for two different blockage scenarios (office and car-park) leads to opposing conclusions about the system performance, and that the correct choice of numerology may improve this performance by as much as hundreds of Mbps. Effectively, we identify conditions under which it may be favourable to use a given numerology, which shall provide insights important for the design of blockage-aware scheduling mechanisms for 5G.

The rest of the paper is organised as follows. 
In Section \ref{sec:related_work}, we present the state-of-the-art and how our work goes beyond it.
In Section \ref{sec:system_model}, we describe our system model.
In Section \ref{sec:performance_metric}, we describe our performance metric.
In Section \ref{sec:numerical_results} we analyse the link performance comparing the numerologies and the blockage conditions.
Finally, we draw our conclusions in Section \ref{sec:conclusion}.
    
    \begin{table}[t]
    \caption{Resource Block Numerology Configuration}
    \centering
    \begin{tabular}{lrrr}
    \toprule
    \multicolumn{1}{c}{\multirow{2}{*}{$\mu$}} & \multicolumn{1}{c}{TTI [ms]} & \multicolumn{1}{c}{CP length [us]} & \multicolumn{1}{c}{Bandwidth [kHz]} \\
    \multicolumn{1}{c}{}    & \multicolumn{1}{c}{$2^{-\mu}$}    & \multicolumn{1}{c}{$4.69/2^\mu$}    & \multicolumn{1}{c}{$2^{\mu}\cdot 15 \cdot 12$} \\ \midrule
    0   & 1         & 4.690     & 180       \\
    1   & 0.5       & 2.345     & 360       \\
    2*   & 0.25      & 1.172     & 720       \\
    3*   & 0.125     & 0.586     & 1440      \\
    4*   & 0.0625    & 0.293     & 2880      \\ \bottomrule
    \multicolumn{4}{l}{\textsubscript{* 5G NR Rel-15 in sub-\unit[6]{GHz} bands can only use $\mu \leq 2$ numerology,}} \\
    \multicolumn{4}{l}{\textsubscript{ while in mmWave bands, only $\mu > 2$ \cite{5gnr_38101-2}.}}
    
    \end{tabular}
    \label{tab:resource_type}
    
    \end{table}

\section{Related Work}
\label{sec:related_work}
The literature on blockage mitigation in mmWave communication is mostly focused on techniques that rely on spatial macro-diversity. Such techniques allow the transmitter to find an alternative physical path for the mmWave signal when the primary \ac{LOS} path fails due to a blockage event.
The main techniques considered are:
(i) \textit{reflectors}: usage of surfaces made of materials that reflect the mmWave signal to cover an obstructed spot through a \ac{NLOS} path \cite{kwon2018multibeam,feng2017dealing}; 
(ii) \textit{relays}: forwarding the transmission to a relay node that has a \ac{LOS} path with the \ac{UE} \cite{wu2017coverage,yang2018sense}; 
(iii) \textit{movable}: moving the \ac{AP} location during the transmission to a position where there is a \ac{LOS} path \cite{bao2017blockage}; 
(iv) \textit{multi-connectivity}: associating the \ac{UE} with multiple \acp{AP}, so the \ac{UE} can have a \ac{LOS} path served by a backup \ac{AP} \cite{tatino2018maximum,petrov2018achieving}.

It is the responsibility of the \ac{MAC} layer to coordinate the extra communication nodes (e.g., relay nodes, neighbour \acp{AP}), and provide a smooth handover between the \acp{AP}, relays, or reflectors when the mmWave signal power fades due to blockage \cite{tesema2017multiconnectivity,giordani2016multiconnectivity,petrov2017dynamic,polese2017improved}.
However, the intermittent blockages together with fixed \ac{TTI} may lead to poor utilisation of the transmission resources.
Therefore, to avoid this under-utilisation, we propose the application of flexible numerology to mitigate blockage effects through \ac{MAC} layer transmission time adaptation.

In state-of-the-art flexible numerology has been applied to improve the network latency where the \ac{TTI} is optimised according to a latency deadline restriction  \cite{ibrahim2016numerology} and according to the traffic pattern \cite{patriciello20185g}.
Also, it has been applied to improve the frame spectral efficiency when multiplexing different types of services, e.g., \ac{eMBB} and \ac{URLLC} \cite{lagen2018subband,you2018resource}.

\section{System Model}
\label{sec:system_model}
\begin{figure}[t]
    \centering
    \begin{subfigure}[b]{.55\linewidth}
    \includegraphics[width=\linewidth]{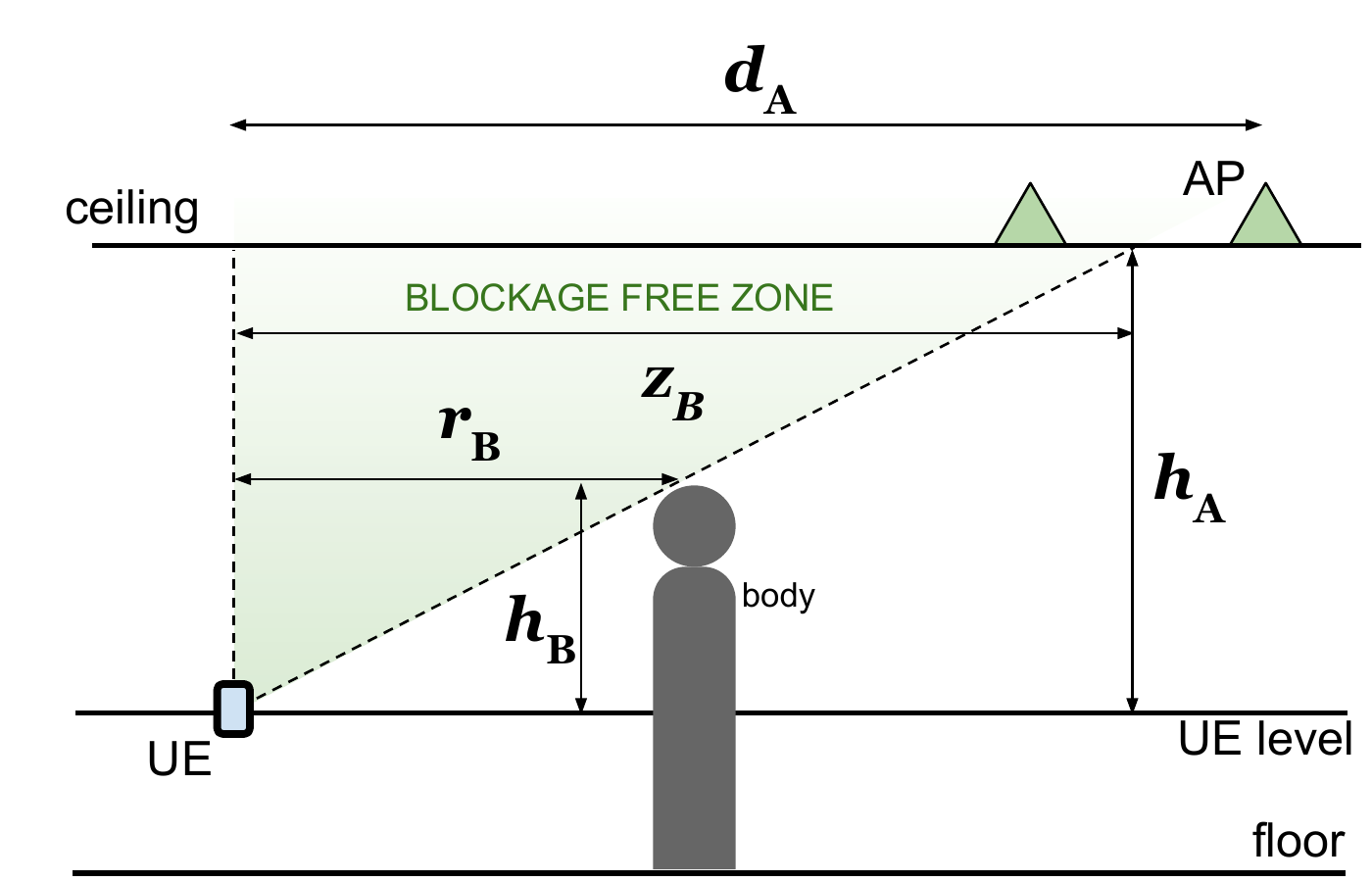}
    \caption{Side view.}
    \label{fig:body_blockage_side_view}
    \end{subfigure}%
    \begin{subfigure}[b]{.45\linewidth}
    \includegraphics[width=\linewidth]{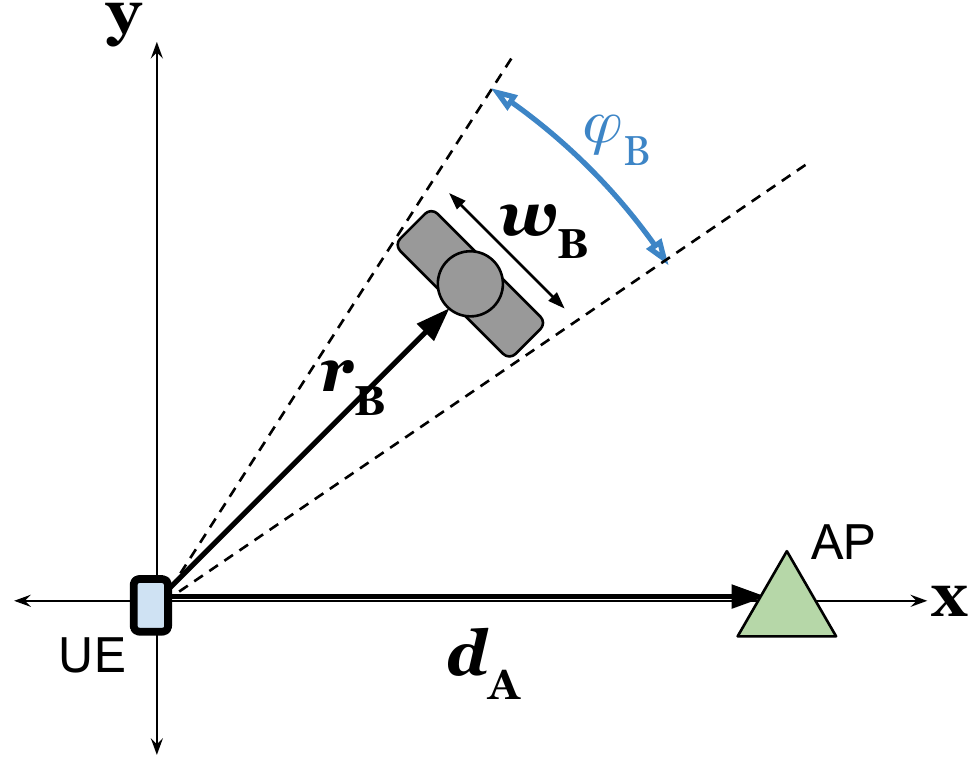}
    \caption{Top view.}
    \label{fig:body_blockage_top_view}
    \end{subfigure}
    \caption{Body blockage model. An \ac{AP} inside the blockage free zone is never blocked by the body regardless of its orientation. (a) For the given body and \ac{AP} heights, an \ac{AP} is inside the blockage free zone when $d_\mathrm{A} < z_\mathrm{B}$. (b) Outside this zone, the \ac{AP} is blocked when its orientation lies in the shadowed cone of width $\varphi_\mathrm{B}$.}
    \label{fig:body_blockage_model}
\end{figure}

We consider a single cell, with an \ac{AP} installed on the ceiling or a lamppost, transmitting a 5G \ac{OFDM} frame to a \ac{UE} at a distance $d_\mathrm{A}$ in the horizontal plane.
The \ac{AP} is installed at a height $h_\mathrm{A}$ above the UE level, as illustrated in Figure \ref{fig:body_blockage_side_view}. This setup shall generalise over the two deployment scenarios considered in \cite{yoo2017channel}, but is also in-line with the 3GPP-defined scenarios for 5G mmWave system evaluation \cite{5gnr_38913}: indoor office with ceiling-mounted access points and outdoor car-park with lamppost mounted access points.

We assume the resource allocation decision in the \ac{AP} is made every \unit[$\tau$]{ms}, which we refer to as the \textit{\ac{SI}}. For ease of exposure, we consider the link performance as experienced by a single user, attached to a single cell. The cell's bandwidth is $b$ and can be filled using flexible numerology $\mu$ with resource blocks of bandwidth $b_\mu$ and \ac{TTI} $t_\mu$, as illustrated in Figure \ref{fig:example_scheduling_5G}.

    \subsection{Blockage Probability Model}
    Whether or not the \ac{LOS} path to the \ac{UE} is blocked during a slot within the \ac{SI} depends on the blockage probability.
    We define the blockage probability as the probability of a slot being blocked during a given interval. This probability is given by the probability of self-body blockage as in \cite{firyaguna2017coverage}:
    \begin{equation}
        p = \left\{\;
        \begin{split}
            & \frac{1}{\pi} \arctan\left(\frac{w_\mathrm{B}}{2 r_\mathrm{B}}\right), \; d_\mathrm{A} \geq z_\mathrm{B}; \\
            & 0, \; \text{otherwise};
        \end{split}
        \; \right.
        \label{eq:prob_blockage}
    \end{equation}
    where $w_\mathrm{B}$ is the body width, $r_\mathrm{B}$ is the distance between the body and the UE, and $h_\mathrm{B}$ is the distance between the UE level and the top of the body, and $z_\mathrm{B}=r_\mathrm{B} \frac{h_\mathrm{A}}{h_\mathrm{B}}$ is the self-body blockage free zone radius, as illustrated in Figure \ref{fig:body_blockage_model}.
    
    \subsection{Signal-to-Noise Ratio}
    For modelling of the mmWave signal propagation, we consider the experimentally-validated channel model proposed in \cite{yoo2017channel}.
    The model consists of the path loss and the composite Gamma-Nakagami-m fading, whose parameters take one of two values depending on whether the user body blocks the \ac{LOS} path, meaning that the model parameters change with the random blockage state. The fading and path-loss coefficients for that model were estimated from the experimental data collected for a mmWave \ac{AP} operating at \unit[60]{GHz} in \cite{yoo2017channel}.
    
    We define the set of the two possible blockage states as $ \chi = \mbox{\footnotesize $\{\mathrm{LOS},\mathrm{NLOS}\}$}$. 
    Hence, given the blockage state $X=x \: \in \chi$, we can define the path loss as $l_x = \ell_x \cdot \left(\sqrt{d_\mathrm{A}^2+h_\mathrm{A}^2}\right)^{-\nu_x}$,
    where $\sqrt{d_\mathrm{A}^2+h_\mathrm{A}^2}$ is the Euclidean distance from the \ac{AP} to the \ac{UE}, $\ell_x$ is the path loss at one metre distance under free space propagation, and $\nu_x$ is the attenuation exponent.
    
    Instead of treating each fading component individually, we consider the fading gain as a single random variable $H_x$ with a composite fading distribution, as obtained in \cite{laourine2009performance}. This approach allows us to define the complementary cumulative distribution function (ccdf) of the SNR in (\ref{eq:snr}) based on the formula in \cite[(15)]{laourine2009performance} as follows:
    \begin{multline}             
        F^c_{Y|X}(y;\:\bar{y}_x,m_x,\alpha_x,\beta_x) \\
        = \mathsf{A} \Gamma(m_x) \sum\limits_{i=0}^{m_x} \frac{2^i y^{m_x-i}}{(BD)^i(m_x-i)!} \frac{\mathcal{K}_{m_x-i+\frac{1}{2}}(\mathsf{B}\sqrt{\mathsf{C}+\mathsf{D}y})}{(\mathsf{B}\sqrt{\mathsf{C}+\mathsf{D}y})^{m_x-i+\frac{1}{2}}}
        \label{eq:ccdf_snr}
    \end{multline}
    where $\bar{y}_x = \frac{\rho}{\sigma} \: l_x$ is the SNR without the fading component, the variable $m_x$ is the Nakagami-m fading parameter, ($\alpha_x,\beta_x$) are the Gamma shadowing parameters, $\mathcal{K}_{o}(\cdot)$ is the modified Bessel function of the second kind of order $o$, and the used constants are:
    \begin{equation}
        \begin{split}
        \mathsf{A} &= \frac{(\alpha_x \Bar{y}_x)^\frac{1+2m_x}{4}}{\Gamma(m_x)} \sqrt{\frac{2\alpha_x\beta_x}{\pi}} \exp(\alpha_x\beta_x) \left(\frac{m_x}{\Bar{y}_x}\right)^{m_x}, \\
        \mathsf{B} &= \beta_x \sqrt{\frac{\alpha_x}{\Bar{y}_x}}, \\
        \mathsf{C} &= \alpha_x \Bar{y}_x, \\
        \mathsf{D} &= 2m_x/\beta_x,
        \end{split}
    \end{equation}
    where $\Gamma(\cdot)$ is the Gamma function.
    
    Thus, we define the signal-to-noise ratio (SNR) $Y$, conditioned on $X=x$, as:
    \begin{equation}
        Y_x = \frac{\rho}{\sigma} \: l_x \: H_x,
        \label{eq:snr}
    \end{equation}
    where $\rho$ is the transmit power, $\sigma$ is the noise power, and $H_x$ is the fading gain.
    
    \subsection{Transmission Efficiency and Slot Aggregation Efficiency}
    
    We define the transmission efficiency $\eta_\mu \in [0,1]$ of a resource block of type $\mu$ as the decrease in the spectral efficiency for shorter \ac{TTI} due to inter-symbol interference caused by shorter \ac{CP} \cite{lahetkangas2014achieving}.
    The longer the \ac{TTI}, the greater the transmission efficiency, i.e., $\eta_i > \eta_j$ for all $i > j$.
    
    We also define the slot aggregation efficiency $\zeta_\mu \in [0,1]$ as the ratio between the number of symbols that are used for data transmission and the total number of symbols.
    
\section{Performance Analysis}
\label{sec:performance_metric}
We evaluate the performance of our system in terms of the expected data rate.
To calculate it, we consider the probability of blockage, the spectral efficiency of the channel, the transmission efficiency of each resource block, and the slot aggregation efficiency.

\begin{figure}[t]
        \centering
        \includegraphics[width=.9\linewidth]{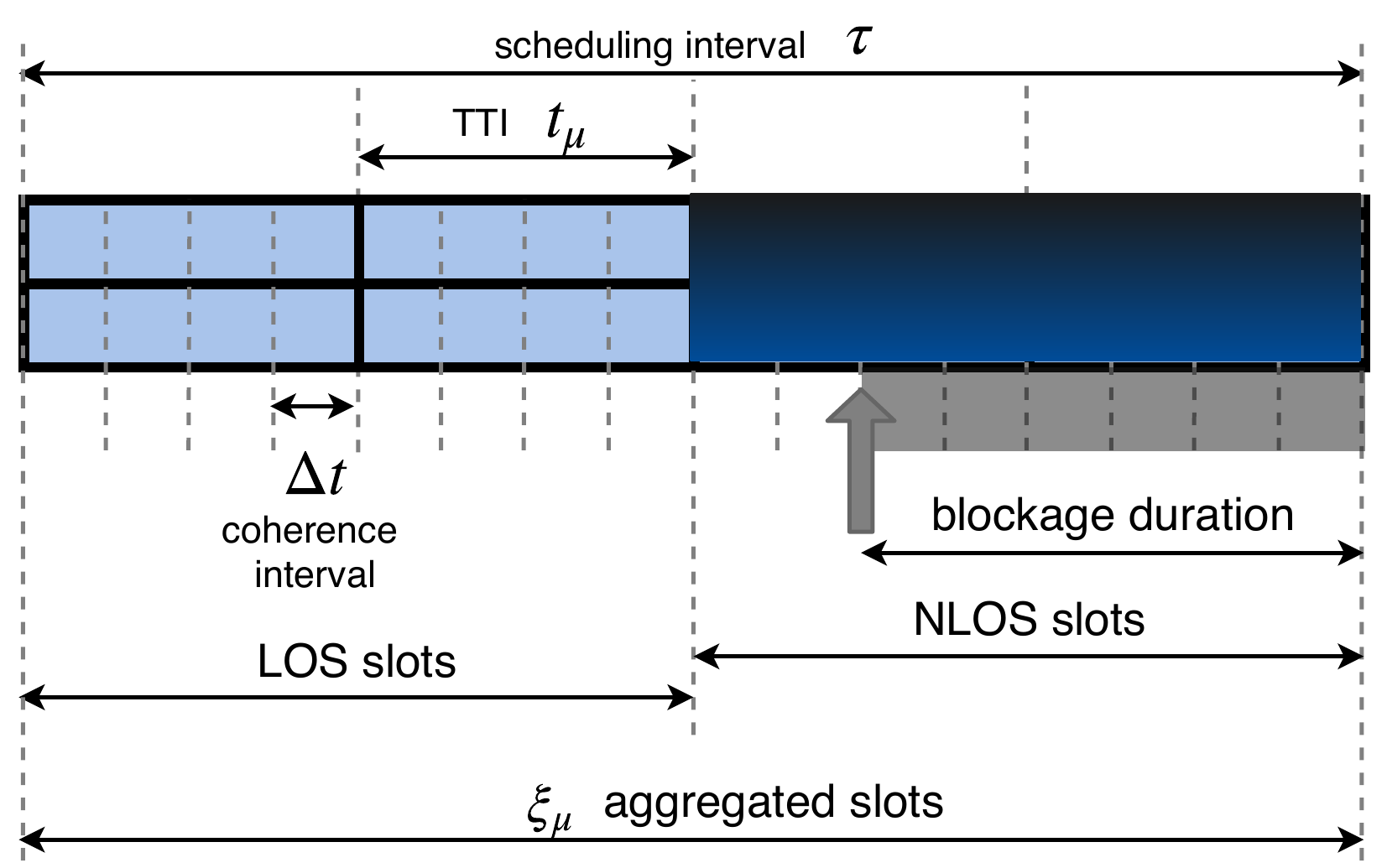}
        \caption{The \ac{SI} $\tau$ contains a sequence of slots with duration $t_\mu$. The \ac{TTI} is a multiple of the blockage \yale{coherence} interval $\Delta t$. When a blockage occurs during one $\Delta t$ interval within a slot, the entire slot and the following ones are considered \ac{NLOS}, as the blockage duration is expected to be significantly larger than the \ac{SI}.}
        \label{fig:time_structure}
    \end{figure}

    We consider that the \ac{TTI} $t_\mu$ is a multiple of $\Delta t$ (see Figure \ref{fig:time_structure}), which is the blockage \yale{coherence} interval. Within this interval, the blockage event has probability $p$ as described in (\ref{eq:prob_blockage}) and is independent of the previous interval, but once the first blockage event happens, we assume that all the following slots within the \ac{SI} $\tau$ are also blocked\footnote{For 5G NR it has been shown in \cite{jain2018driven} that the body blockage duration can be in the order of \unit[100]{ms} (due to low mobility of pedestrians) versus the \unit[10]{ms} duration frames.}.
    Consequently, a slot, that contains $k$ coherence intervals ($t_\mu = k \cdot \Delta t$), is transmitted in \ac{LOS} if no blockage has occurred in previous slots and in each of its own coherence intervals. Thus, the probability of the $i$-th slot being in \ac{LOS} is:
    \begin{equation}
    \begin{split}
        \Pb[X_i = \mbox{\scriptsize $\mathrm{LOS}$}] = (1-p)^{k \cdot i} = (1-p)^{i \frac{t_\mu}{\Delta t}}.
        \label{eq:prob_los}
    \end{split}
    \end{equation}
    
    The spectral efficiency $S$ can be expressed as $S=\log(1+Y)$, and has only non-negative values. Hence, the expected value conditioned to the blockage state of the $i$-th slot can be represented by this integral which can be efficiently computed numerically:
    \begin{equation}
        \Eb[S_{x_i}|X_i=x_i] = \int\limits_0^\infty  F_{Y|X}^c(2^s-1; \: \bar{y}_x,m_x,\alpha_x,\beta_x) \: \dd{s},
        \label{eq:expected_se}
    \end{equation}
    where $F_{Y|X}^c(\cdot)$ is ccdf of the SNR defined in (\ref{eq:ccdf_snr}).
    Thus, the expected spectral efficiency with respect to the blockage state of the $i$-th slot is:
    \begin{equation}
    \begin{split}
        \Eb_{X_i,S}\left[S_{X_i}\right] &= \sum\limits_{x_i \in \chi} \Pb[X_i=x_i] \Eb[S_{x_i}|X_i=x_i] \\
        &\overset{(a)}= (1-p)^{i \frac{t_\mu}{\Delta t}} \cdot  \Eb[S_\mathrm{LOS}|X=\mbox{\scriptsize $\mathrm{LOS}$}] \\
        &+ \big(1-(1-p)^{i \frac{t_\mu}{\Delta t}}\big) \cdot  \Eb[S_\mathrm{NLOS}|X=\mbox{\scriptsize $\mathrm{NLOS}$}],
        \label{eq:expected_se_x}
    \end{split}
    \end{equation}
    where (a) comes from using (\ref{eq:prob_blockage}) and from the fact that the channel is constant with the same blockage state.
    
    Finally, the expected data rate using the numerology $\mu$ ($\bar{R_\mu}$) of a slot aggregation of $\xi_\mu$ slots is given by the expectation of the sum of the spectral efficiency of each slot multiplied by the frame bandwidth and the transmission and slot aggregation efficiencies:
    \begin{equation}
        \bar{R_\mu} = b \cdot \zeta_\mu \cdot \eta_\mu \cdot  \Eb_{X,S}\left[\sum\limits_{i=1}^{\xi_\mu} S_{X_i}\right].
    \end{equation}
    Then, using (\ref{eq:expected_se_x}), $\bar{R_\mu}$ becomes as expressed in (\ref{eq:expected_rate}):
    \begin{figure*}
    \begin{equation}
        \begin{split}
        \bar{R_\mu} &= b \, \zeta_\mu \, \eta_\mu \, \sum\limits_{i=1}^{\xi_\mu} \Eb_{X,S}\left[S_{X_i}\right] 
        = b \, \zeta_\mu \, \eta_\mu \, \sum\limits_{i=1}^{\xi_\mu} \left( (1-p)^{i \frac{t_\mu}{\Delta t}} \,  \Eb[S_\mathrm{LOS}|X=\mbox{\scriptsize $\mathrm{LOS}$}] + \big(1-(1-p)^{i \frac{t_\mu}{\Delta t}}\big) \, \Eb[S_\mathrm{NLOS}|X=\mbox{\scriptsize $\mathrm{NLOS}$}] \right) \\
        &= b \, \zeta_\mu \, \eta_\mu \, \Big( \xi_\mu \,  \Eb[S_\mathrm{NLOS}|X=\mbox{\scriptsize $\mathrm{NLOS}$}] + 
        \frac{(1-p)^{ \frac{t_\mu}{\Delta t}}((1-p)^{\xi_\mu \frac{t_\mu}{\Delta t}}-1)}{(1-p)^{ \frac{t_\mu}{\Delta t}}-1} \left( \Eb[S_\mathrm{LOS}|X=\mbox{\scriptsize $\mathrm{LOS}$}] - \Eb[S_\mathrm{NLOS}|X=\mbox{\scriptsize $\mathrm{NLOS}$}] \right) \Big)
    \label{eq:expected_rate}
        \end{split}
    \end{equation}
    \end{figure*}

\section{Numerical Results}
\label{sec:numerical_results}
In this section, we show the benefits of the application of the flexible numerology to blockage mitigation in 5G NR mmWave systems.
We compare the results for three types of resource blocks supported by mmWave NR (i.e., $\mu=2,3,4$ according to 5G NR Rel-15 \cite{5gnr_38211}), under different blockage scenarios.

We consider two blockage scenarios: a UE held in a pocket (\unit[$r_\mathrm{B}=0$]{cm}, \textit{UE in pocket}), and a UE operated with the hand (\unit[$r_\mathrm{B}=30$]{cm}, \textit{UE in hand}). The \textit{UE in pocket} scenario is a severe blockage condition where the body is obstructing half of the \yale{angle-of-view}, thus, the probability of blockage is $p=0.5$.
The \textit{UE in hand} scenario is a common blockage condition where the user is, for example, operating an app in the mobile phone, and the body obstructs a smaller angle than in the \textit{UE in pocket} scenario.
We consider two environments: an indoor open \textit{office} and an outdoor \textit{car park}. This setup reflects the scenarios and environments characterised in \cite{yoo2017channel}, for which coefficients of the fading and path loss models we use were estimated. These coefficients are listed on the left side of Table \ref{tab:model_parameters} (based on \cite[Table I]{yoo2017channel}).

We assume that the transmission efficiency $\eta_\mu$ decreases by 5\% with each increment in the numerology $\mu$ (i.e., $\eta_2=1.00$, $\eta_3 = 0.95$ and $\eta_4 = 0.90$). Accurate values for the numerology-dependent transmission efficiency can be obtained following the calculations presented in \cite{lahetkangas2014achieving}.
We assume that the slot structure is as described in \cite{mogensen2013small}, where each slot consists of 14 symbols. In a slot aggregation of $\xi_\mu = \tau/t_\mu$ slots, the two first symbols are used for downlink and uplink control, the third is used for demodulation reference signal, and the rest of the symbols is for data. Thus, the slot aggregation efficiency is $\zeta_\mu = 1 - \frac{3 t_\mu}{14 \tau}$.
We set the frame bandwidth as \unit[100]{MHz} \cite{5gnr_38211} and we evaluate the performance with the \ac{SI} \unit[$\tau=1$]{ms} (unless specified otherwise), as in the legacy LTE scheduling.
For our analysis, we set \unit[$\Delta t=0.0625$]{ms} as the shortest \ac{TTI} among the numerologies considered.
All other fixed system parameters are shown in Table \ref{tab:model_parameters} (right side).
\begin{table*}[]
\centering
\caption{Model Parameters}
\label{tab:model_parameters}
\begin{tabular}{lcrcr|crcr|cc|lcr}
\toprule
\multicolumn{11}{c|}{Channel Model} & \multicolumn{3}{c}{System Model} \\
\hline
\multirow{3}{*}{Environment} & \multicolumn{4}{|c|}{Path Loss} & \multicolumn{4}{c|}{Shadowing} & \multicolumn{2}{c|}{Small-Scale Fading} & Transmit Power & $\rho$ & 20 dBm\\  
 & \multicolumn{2}{|c}{LOS} & \multicolumn{2}{c|}{NLOS} & \multicolumn{2}{c}{LOS} & \multicolumn{2}{c|}{NLOS} & \multicolumn{1}{c}{LOS} & \multicolumn{1}{c|}{NLOS} & Noise Density & $\sigma/b$ & -174 dBm/Hz \\ 
 & \multicolumn{1}{|c}{$\nu$} & \multicolumn{1}{c}{$\ell$ (dB)} & $\nu$ & \multicolumn{1}{c|}{$\ell$ (dB)} & $\alpha$ & \multicolumn{1}{c}{$\beta$} & $\alpha$ & \multicolumn{1}{c|}{$\beta$} & \multicolumn{1}{c}{$m$} & \multicolumn{1}{c|}{$m$} & Body Width & $w_\mathrm{B}$ & 40 cm \\
Office & \multicolumn{1}{|c}{1.18} & 45.1 & \multicolumn{1}{c}{1.07} & 57.4 & \multicolumn{1}{c}{7.01} & 0.15 & \multicolumn{1}{c}{5.77} & 0.20 & \multicolumn{1}{c}{2.64} & \multicolumn{1}{c|}{2.35} & Body Height* & $h_\mathrm{B}$ & 40 cm \\
Car Park & \multicolumn{1}{|c}{1.53} & 48.7 & \multicolumn{1}{c}{1.98} & 88.8 & \multicolumn{1}{c}{10.30} & 0.11 & \multicolumn{1}{c}{5.11} & 0.23 & \multicolumn{1}{c}{8.50} & \multicolumn{1}{c|}{2.74} & AP Height* & $h_\mathrm{A}$ & 5 m \\ 
\bottomrule
\multicolumn{14}{r}{\textsubscript{* with respect to the UE level}}
\end{tabular}
\end{table*}

\subsection{Environment and Blockage Impact}
\begin{figure}[]
    \centering
    \begin{subfigure}{.9\linewidth}
    \includegraphics[width=\linewidth,trim={0cm 0 0 0},clip]{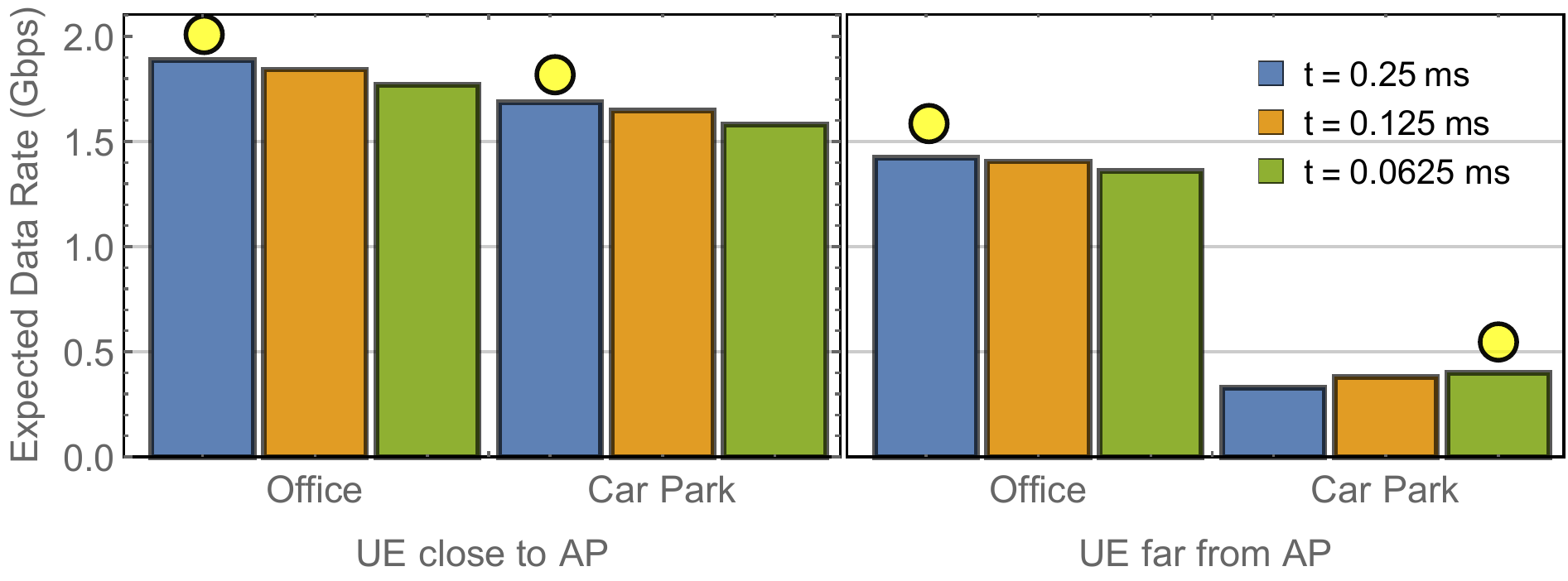}
    \caption{\textit{UE in hand} scenario.}
    \label{fig:avgR_bar_hand_ta1ms}
    \end{subfigure}
\hfill
    \begin{subfigure}{.9\linewidth}
    \centering
    \includegraphics[width=\linewidth,trim={0.0cm 0 0 0},clip]{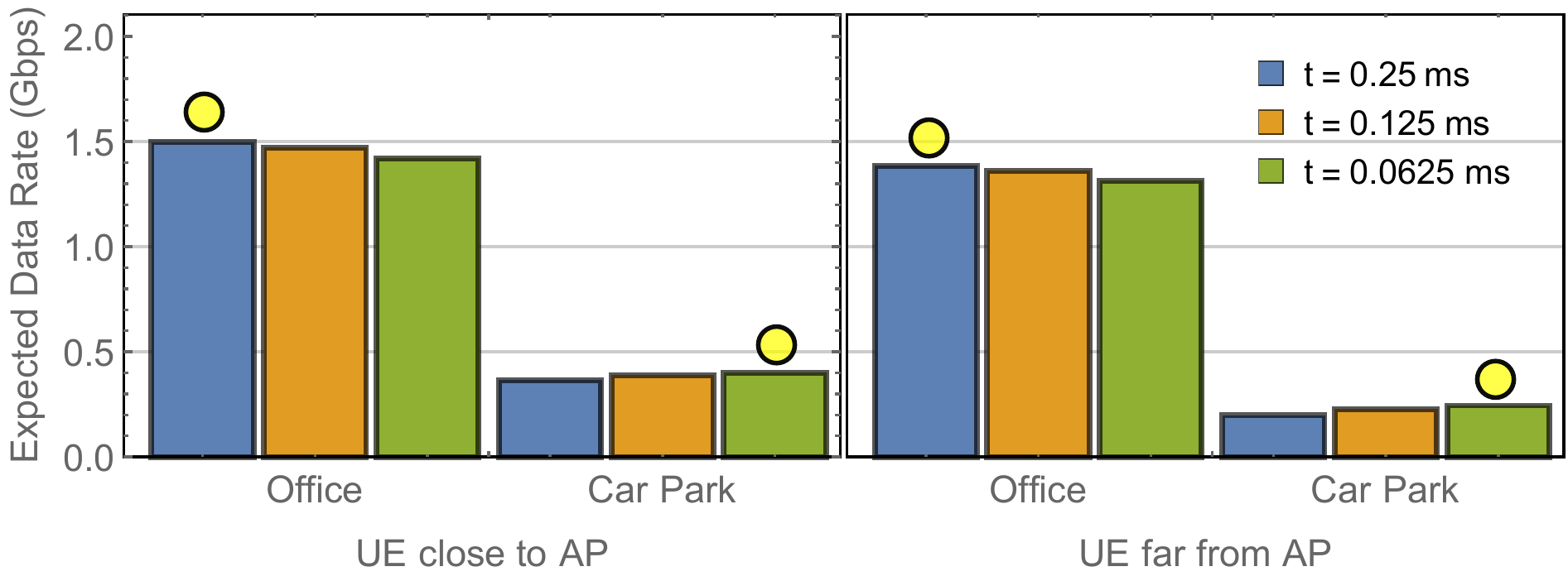}
    \caption{\textit{UE in pocket} scenario.}
    \label{fig:avgR_bar_pocket_ta1ms}
    \end{subfigure}
    \caption{Expected data rate using different \ac{TTI}s with \ac{SI} \unit[$\tau=1$]{ms}, in two environments (\textit{office} and \textit{car park}), at two distances between the UE and AP (\unit[1]{m} and \unit[10]{m}), and in two blockage scenarios (\textit{UE in hand} and \textit{UE in pocket}). The yellow disks indicates the recommended numerology for scheduling in the given scenario.}
    \label{fig:avgR_bar_environment}
\end{figure}

In this subsection, we evaluate the impact of two environments (an \textit{office} and a \textit{car park}), with distinct channel characteristics, and two blockage scenarios (\textit{UE in hand} and \textit{UE in pocket}) on the expected data rate of mmWave communication. 
We also compare the scenarios where the UE is close to the AP (\unit[$d_\mathrm{A}=1$]{m}) and where the UE is far from the AP (\unit[$d_\mathrm{A}=10$]{m}).
The results are shown in Figure \ref{fig:avgR_bar_environment}.

From the left side of Figure \ref{fig:avgR_bar_hand_ta1ms}, we see that the expected data rate is higher when using a \ac{TTI} of \unit[$t_2=0.25$]{ms} (blue bar), compared to other \ac{TTI}s, for a user close to the AP and operating the UE with the hand. 
In this case, the blockage probability is very low, allowing for high transmission efficiency of the resource block with long slot duration to have a more significant impact on the data rate than the blockage. 
On the right-hand side of that same figure, we observe a decrease in the expected data rate because of the dual-effect of increased path loss and blockage probability for a user further away from the AP.
The \textit{car park} environment suffers more from blockages compared to the \textit{office} environment as there is less power in the \ac{NLOS} signal, likely due to lack of reflecting/scattering environment.
Thus, short \ac{TTI} (\unit[$t_4=0.0625$]{ms}, green bar) mitigates the blockage effects by increasing the expected number of slots in LOS and, then, yields better performance in the \textit{car park} environment.

From Figure \ref{fig:avgR_bar_pocket_ta1ms}, we note that the expected data rate has similar trends when we vary the distance between the UE and the AP, as the considered range of distances have little effect on the blockage probability for the UE in the pocket. 
We see that the user in the \textit{office} environment achieves highest expected data rate using \ac{TTI} \unit[$t_2=0.25$]{ms}, and in the \textit{car park} environment, the user achieves highest expected data rate using \ac{TTI} \unit[$t_4=0.0625$]{ms} in both cases (UE close to AP and far from AP) of the \textit{UE in pocket} scenario.

\subsection{Scheduling Interval Impact}
\begin{figure}[]
    \centering
    \begin{subfigure}{.9\linewidth}
    \includegraphics[width=\linewidth,trim={0cm 0 0 0},clip]{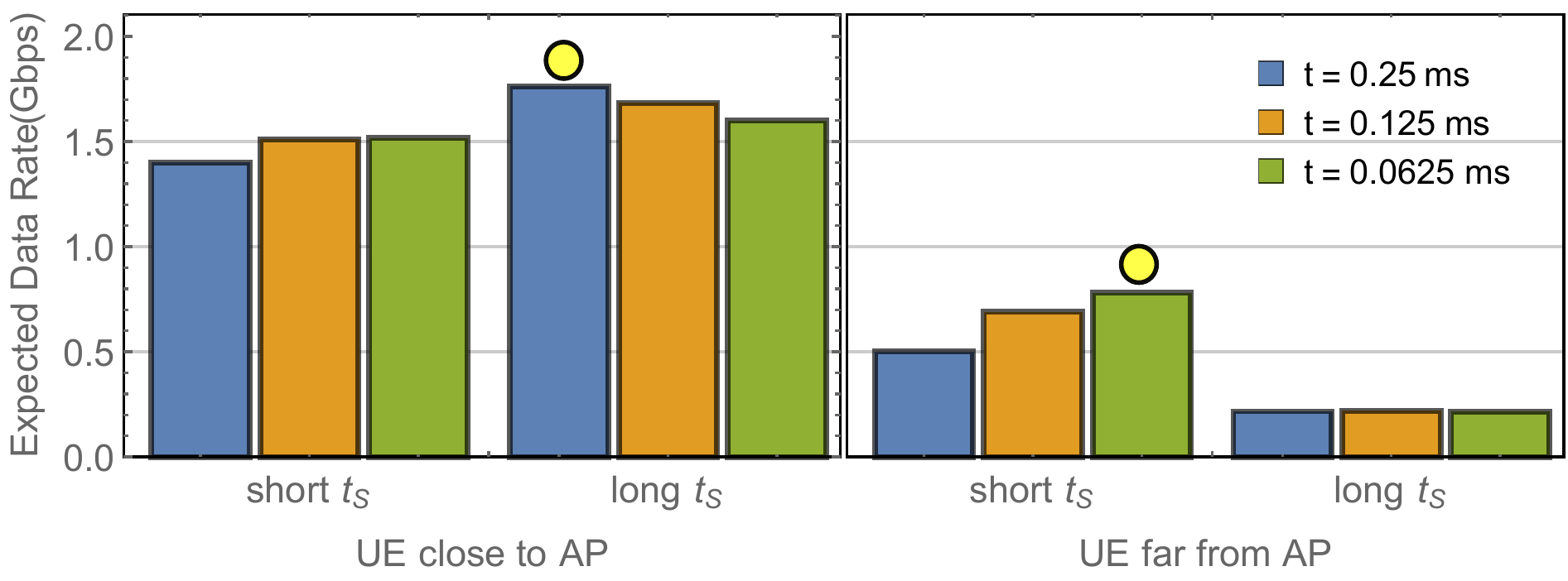}
    \caption{\textit{UE in hand} scenario.}
    \label{fig:avgR_bar_carpark_hand}
    \end{subfigure}
\hfill
    \begin{subfigure}{.9\linewidth}
    \centering
    \includegraphics[width=\linewidth,trim={0cm 0 0 0},clip]{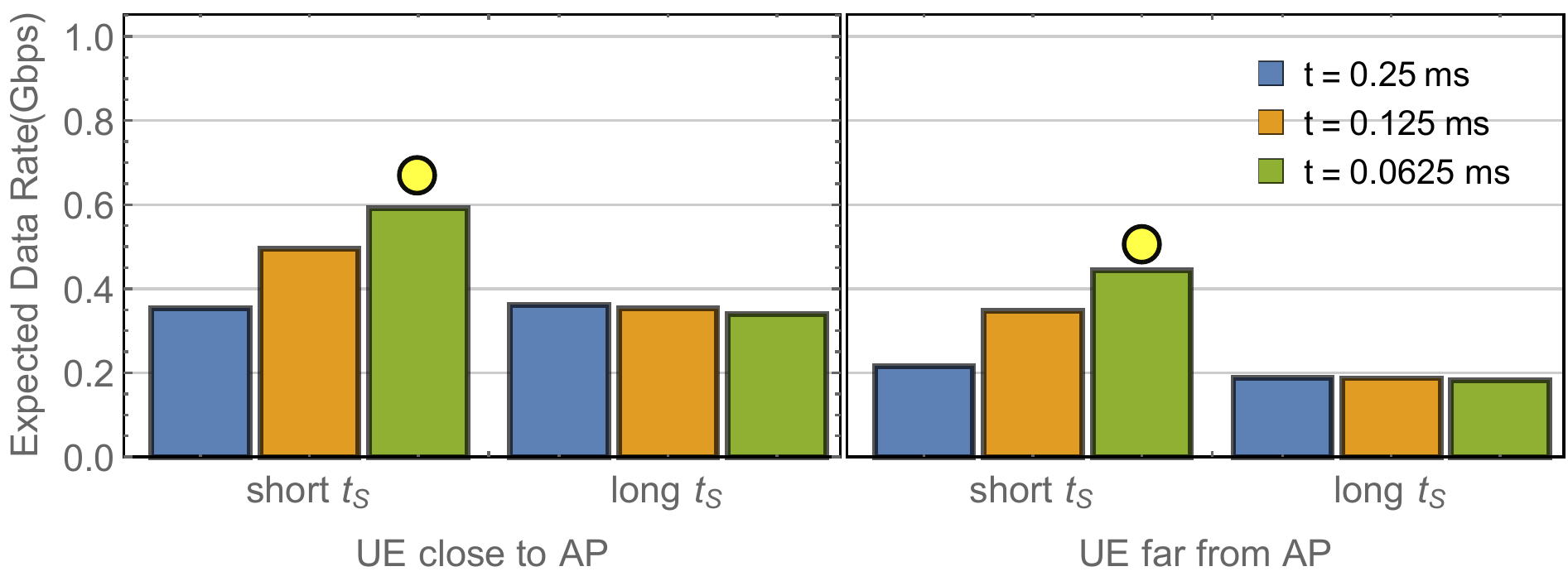}
    \caption{\textit{UE in pocket} scenario.}
    \label{fig:avgR_bar_carpark_pocket}
    \end{subfigure}
    \caption{Expected data rate using different \ac{TTI}s, with two \ac{SI}s (\unit[$\tau=0.25$]{ms}, \unit[$\tau=5$]{ms}), in the \textit{car park} environment, at two distances between the UE and AP (\unit[1]{m} and \unit[10]{m}), and in two blockage scenarios (\textit{UE in hand} and \textit{UE in pocket}). The yellow disks indicates the recommended numerology for scheduling in the given scenario.}
    \label{fig:avgR_bar_scheduling}
\end{figure}

In LTE networks, the scheduler makes the allocation decision every \ac{TTI}, which has a fixed duration of \unit[1]{ms}, and the decision is valid until the next \ac{TTI} \cite{capozzi2013downlink}. 
The 5G NR allows slot aggregation, in which the aggregation duration can span two or more slots to reduce control overhead.
Hence, the scheduling decision interval is no longer fixed and can vary with the slot aggregation size.
Here, we evaluate the impact of the aggregation overhead reduction by comparing the \ac{SI}s \unit[$\tau=0.25$]{ms} (short \ac{SI})\footnote{A \ac{SI} of \unit[$\tau=0.25$]{ms} allows to allocate at least one of the longest \ac{TTI} (\unit[$t_4=0.25$]{ms}) considered.} and \unit[$\tau=5$]{ms} (long \ac{SI})\footnote{Any longer \ac{SI} between \unit[5 and 10]{ms} leads to similar results.}. The results considering the \textit{car park} environment are shown in Figure \ref{fig:avgR_bar_scheduling}.

From Figure \ref{fig:avgR_bar_scheduling}, we observe that, in most cases, a short \ac{SI} achieves better performance with short \ac{TTI} (green bar) compared to long \ac{TTI} (blue bar). 
The only exception, as we see from the left side of Figure \ref{fig:avgR_bar_carpark_hand}, is the scenario of a user close to the AP operating the UE with the hand.
In this scenario, the best performance is achieved by long \ac{SI} with long \ac{TTI}.
This is because the effects of increased transmission efficiency (long \ac{TTI}) and reduced overhead (long \ac{SI}) have a higher impact on the expected data rate than the low blockage probability.
In other cases, i.e., UE far from AP in \textit{UE in hand} scenario and both cases in \textit{UE in pocket} scenario (see Figure \ref{fig:avgR_bar_carpark_pocket}), the blockage probability is high and has more impact on the expected data rate than the overhead reduction.
Thus, there is no benefit in increasing the \ac{SI} in those cases.

To sum up, the appropriate combination of numerology and scheduling interval is essential to achieve the best performance. For example, in the cases where the blockage has more significant impact on the expected data rate (e.g., UE far from AP or UE in pocket), the use of short \ac{SI} with short \ac{TTI} is recommended to avoid prolonged exposure to blockage interruptions.
On the other hand, in the cases where the blockage has less impact (e.g., UE close to AP), the use of long \ac{SI} with long \ac{TTI} is recommended to take advantage of the reduced overhead and the high transmission efficiency.

\section{Conclusion}
\label{sec:conclusion}
In this paper, we proposed that 5G NR flexible numerology be used to mitigate the negative effects of body blockage in 5G \ac{mmWave} systems. We presented an analytical framework that allowed us to show and understand the benefits of our proposed application of flexible numerology. We showed that different blockage conditions require different combinations of numerology and slot aggregation to achieve the best performance, as presented in Table \ref{tab:recommendation}.
The effectiveness of flexible numerology in mmWave blockage scenarios is a consequence of the trade-off between the high transmission efficiency, achieved with long \acp{TTI}, and the high probability of LOS transmission, achieved using short \acp{TTI}.
\begin{table}[b]
\centering
\caption{Recommended Numerology $\mu$ and Scheduling Interval $\tau$ (in \unit[]{ms})}
\label{tab:recommendation}
\begin{tabular}{c|c|c|c|c|}
\toprule
 \multicolumn{1}{|c|}{} & \multicolumn{2}{c|}{Office} & \multicolumn{2}{c|}{Car Park} \\ \cline{2-5} 
 \multicolumn{1}{|c|}{\textbf{Environment}} & UE close & UE far & UE close & UE far \\
 \multicolumn{1}{|c|}{} & to AP & from AP & to AP & from AP \\ \hline
\multicolumn{1}{|c|}{} & \cellcolor[HTML]{5E82B5}$\mu=2$ & \cellcolor[HTML]{5E82B5}$\mu=2$ & \cellcolor[HTML]{5E82B5}$\mu=2$ & \cellcolor[HTML]{8EAF30}$\mu=4$ \\
\multicolumn{1}{|c|}{\multirow{-2}{*}{UE in hand}} & \cellcolor[HTML]{5E82B5}$\tau=5$ & \cellcolor[HTML]{5E82B5}$\tau=5$ & \cellcolor[HTML]{5E82B5}$\tau=5$ & \cellcolor[HTML]{8EAF30}$\tau=0.25$ \\ \hline
\multicolumn{1}{|c|}{} & \cellcolor[HTML]{5E82B5}$\mu=2$ & \cellcolor[HTML]{5E82B5}$\mu=2$ & \cellcolor[HTML]{8EAF30}$\mu=4$ & \cellcolor[HTML]{8EAF30}$\mu=4$ \\
\multicolumn{1}{|c|}{\multirow{-2}{*}{UE in pocket}} & \cellcolor[HTML]{5E82B5}$\tau=5$ & \cellcolor[HTML]{5E82B5}$\tau=5$ & \cellcolor[HTML]{8EAF30}$\tau=0.25$ & \cellcolor[HTML]{8EAF30}$\tau=0.25$ \\  \bottomrule
\end{tabular}
\end{table}

This work is a stepping stone to further studies on the application of flexible numerology to blockage mitigation in 5G-mmWave networks. Further work is needed to investigate the implications of multiple users or services sharing the frame, as well as how the effectiveness of flexible numerology affects the blockage mitigation via macro-diversity. Nonetheless, the results we have shown thus far should motivate the development of new scheduling algorithms/policies for the 5G NR mmWave frame.

\appendices

\section*{Acknowledgement}

This publication has emanated from the research conducted within the scope of \textit{NEMO (Enabling Cellular Networks to Exploit Millimetre-wave Opportunities)} project financially supported by the Science Foundation Ireland (SFI) under Grant No. 14/US/I3110 and with partial support of the European Regional Development Fund under Grant No. 13/RC/2077.
We are thankful to Danny Finn who kindly assisted us with the Wolfram Mathematica\textsuperscript{\textregistered} software.

\section{}
\label{ap:scripts}
All scripts used to generate the presented results were written in Wolfram Mathematica\textsuperscript{\textregistered} and are available in \textit{https://github.com/firyaguna/wolfram-flexible-numerology}.

\bibliographystyle{IEEEtran}
\bibliography{myreferences}

\end{document}